\documentclass{ws-procs9x6}

\usepackage{bm}
\DeclareBoldMathCommand{\bfzeta}{\zeta}
\DeclareBoldMathCommand{\bfpi}{\pi}

\renewcommand{\imath}[0]{\mathrm{i}}

\newcommand{\im}[1]{\mathrm{Im}\left.#1\right.}

\begin{document}

\title{Mode contributions to the Casimir effect}

\author{F. INTRAVAIA$^*$\footnote[4]{Present 
address: Theoretical Division, MS B213
Los Alamos National Laboratory, 
Los Alamos NM 87545, U.\ S.\ A.} 
and C. HENKEL}

\address{Institut f\"ur Physik und Astronomie,
Universit\"at Potsdam,\\ 
Karl-Liebknecht-Str. 24/25,
14476 Potsdam, Germany\\
$^*$E-mail: francesco.intravaia@qipc.org\\
www.quantum.physik.uni-potsdam.de}
\begin{abstract}
Applying a sum-over-modes approach to the Casimir interaction between 
two plates with finite conductivity, we isolate and study the contributions of 
surface plasmons and Foucault (eddy current) modes.
We show in particular that  for the TE-polarization eddy currents provide a repulsive force that
cancels, at high temperatures, the Casimir free energy calculated with
the plasma model 
\end{abstract}


\bodymatter

\section{Introduction}\label{intro}
Intense theoretical effort is currently devoted to the understanding of the Casimir
effect for real experimental setups. This involves the impact of temperature, 
finite conductivity, engineered materials,  
and may identify routes to \emph{design} the final Casimir pressure.
Almost all analyses rely on the Lifshitz formula \cite{Lifshitz56, Klimchitskaya09}
where the physical properties of the material are encoded in the scattering
amplitudes (i.e., reflection coefficients in planar geometries). 
Their evaluation at imaginary frequencies obscures, however, how the material
objects modify the modes of the electromagnetic field. 
A `sum over modes' approach is nevertheless possible, even if the 
eigenfrequencies
$\omega_{m}$ are complex (due to material absorption, for example).
For two objects at distance $L$ the Casimir energy at zero temperature can 
be written as \cite{Intravaia08}
\begin{equation}
\label{eq:Casimir-Diss}
E = \frac{\hbar}{2} 
\sideset{}{'}\sum_{p,\mathbf{k}} {\rm Re}\,\Big[\sum_{m}
\big(\omega_{m}-
\frac{2\imath\omega_{m}}{\pi}
\ln\frac{\omega_{m}}{\Lambda}
\big)\Big]^{L}_{\infty}
, \qquad
\im{\Big[\sideset{}{'}\sum_{p,\mathbf{k},m}
\omega_{m}\Big]_{\infty}^{L}}=0
\end{equation}
where the prime indicates that purely imaginary eigenfrequencies are weighted 
with $1/2$. Eq.(\ref{eq:Casimir-Diss}) generalizes Casimir's formula for the 
vacuum energy between two perfect reflectors \cite{Casimir48} and is valid for generic 
(causal) mirrors with arbitrary thickness. Note that one does not simply take 
real parts of the complex eigenfrequencies, as suggested some time 
ago\cite{Langbein70} (see also Ref.\refcite{Sernelius06}).
The logarithmic correction in Eq.(\ref{eq:Casimir-Diss}) is consistent with
the `system+bath' paradigm that describes the thermodynamics
of quantum dissipative systems\cite{Weiss08}. 
In this context, the frequency scale $\Lambda$ is interpreted as the cutoff
frequency of the bath spectral density. The Casimir energy does not depend 
on this constant because of the sum rule in \eqref{eq:Casimir-Diss}.

The sum-over-modes approach 
provides an `anatomic view' of the Casimir effect where contributions from
different modes are clearly identified. This is useful to understand
unusual behaviours and may suggest new ways to 
taylor the Casimir force\cite{Intravaia05,Intravaia07,Intravaia09}.
In the following, we illustrate Eq.\eqref{eq:Casimir-Diss} with the help of 
a few examples.

\section{Dissipative Plasmons at short distance}

\begin{figure}[htp] 
   \centering
   \includegraphics[width=\textwidth]{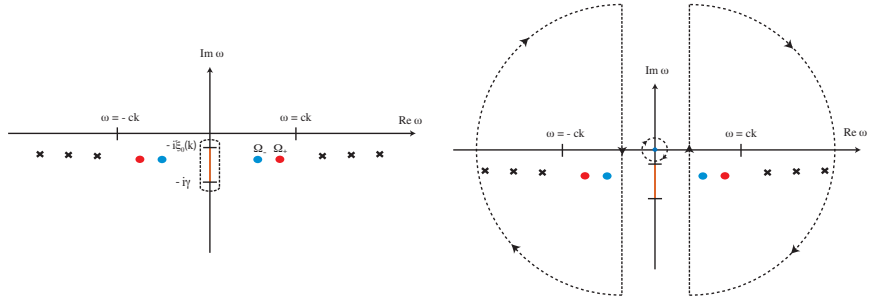} 
\caption{(Left) Complex eigenfrequencies in the parallel plate geometry, for a fixed
wavevector $k$ (not to scale). 
Red and blue points: dissipative surface plasmons. 
Red line: bulk continuum of eddy currents. Black crosses: propagating
modes in the cavity between the plates.
(Right)
A counter-clockwise path around the eddy current continuum is equivalent 
to a clockwise path around the whole complex plane, encircling all other 
modes.}
   \label{ChangePath}
   \label{BranchCut}
\end{figure}

One of the most interesting contributions to the Casimir force originates from 
surface modes bound to the vacuum/medium interface\cite{Barton79}. These
modes have a dispersion relation that splits in two branches, 
$\omega = \Omega_\pm( k )$, as two surfaces are approached. 
Substituting these frequencies in Eq.\eqref{eq:Casimir-Diss}, we get a
plasmonic contribution to the Casimir energy ($A$: surface area)
\begin{equation}
E_{\rm pl} = \frac{\hbar A }{2} 
\int\!\frac{ k {\rm d}k }{ 2\pi } 
{\rm Re}\,\Big[\sum_{i=\pm}
\big(\Omega_{i}(k) -
\frac{2\imath\Omega_{i}(k)}{\pi}
\ln\frac{\Omega_{i}(k)}{\Lambda}
\big)\Big]^{L}_{\infty}
\label{smalldiss1}
\end{equation}
Consider the case of two metals
at a distance smaller than the plasma wavelength 
$\lambda_{\rm pl} = 2\pi c/\omega_{\rm pl}$. We are then in the
quasi-electrostatic regime, and the surface plasmon modes are given 
by\cite{Economou69} (red and blue points in Fig.\ref{BranchCut})
%
\begin{equation}
\Omega_{\pm} = 
	\sqrt{\omega^2_{\pm}-\frac{\gamma^2}{4}}
	- \imath\frac{\gamma}{2}
,\qquad 
\omega^2_{\pm} = \frac{ \omega^2_{\rm pl} }{ 2 }
\left(1\pm e^{ -kL}\right)
\end{equation} 
where $\gamma$ is the damping rate in a Drude description of the metal.
%
%
%
One can easily check that the sum rule 
in Eq.\eqref{eq:Casimir-Diss} is automatically satisfied. 
To leading order in $\gamma\ll \omega_{\rm pl}$ (good conductors) Eq.\eqref{smalldiss1} yields
\begin{equation}
	E_{\rm pl} \approx 
	-\frac{\pi^2 \hbar c A}{720 L^3}
\frac{3}{2}\left(\alpha \frac{L}{\lambda_{\rm pl}}
-
\frac{15\zeta(3)}{\pi^4} 
\frac{ \gamma L}{ c }
\right)
,\qquad
\alpha=1.193\ldots
	\label{eq:short-distance-expansion}
\end{equation}
where $\zeta(3) \approx 1.202$ is a Zeta function.
This corresponds exactly to the total Casimir force calculated in 
Ref.\refcite{Henkel04}, including the dissipative correction.
In fact, in this short distance limit, the Casimir energy 
is completely dominated by the plasmonic 
contribution\cite{Kampen68,Gerlach71,Henkel04}. 
Eq.\eqref{smalldiss1} is valid 
also beyond the good conductor limit, however, and could be used, e.g., to
analyze semiconductors where surface plasmons appear in a different
frequency range and can have much stronger damping.

\section{Eddy currents}

As a second example, consider the contribution from eddy current modes.  
They are connected with low-frequency currents that satisfy a diffusion
equation in the conducting metal\cite{Jackson75} and are completely
absent within the lossless description of the so-called plasma 
model\cite{Klimchitskaya09}. 
We have analyzed these 
modes recently\cite{Intravaia09} and constructed from the `system+bath'
paradigm their quantum thermodynamics. They behave like free Brownian
particles, since the eigenfrequencies of bulk eddy currents are purely
imaginary $\omega_{m}= - \imath \xi_{m}$ ($\xi_{m}>0$). From 
Eq.\eqref{eq:Casimir-Diss}, we get the Casimir energy
\begin{equation}
\label{eq:Eddy}
E_{\rm eddy} = -\sum_{p,\mathbf{k}} \,\Big[\sum_{m}
\frac{\hbar\xi_{m}}{2\pi}
\ln\frac{\xi_{m}}{\Lambda}
\Big]^{L}_{\infty}
\end{equation}
For these modes alone,
the sum rule [Eq.\eqref{eq:Casimir-Diss}] is not satisfied, and the 
eddy current contribution to the Casimir energy depends on the cutoff 
$\Lambda$. This is also well-known from quantum Brownian motion where
bath modes up to $\Lambda$ are entangled to the particle.

Mathematically, eddy currents form a mode continuum that can be identified
in the complex frequency plane from the branch cut of the root 
$k_{m}=\sqrt{\epsilon(\omega)\omega^{2}/c^{2} - k^{2}}$
which describes the propagation of the electromagnetic field inside the 
medium. For a Drude metal, the cut is located between 
$\omega_{m} = -\imath\xi_{0}(\mathbf{k})
\approx -\imath D k^{2}$ (for $k \ll \omega_{\rm pl} / c$)
%
%
and $\omega_{m} = -\imath\gamma$ (see Fig. \ref{BranchCut}),
%
where $D=\gamma(\lambda_{\rm pl}/2\pi)^2$ is the electromagnetic
diffusion constant. 
We get the $L$-dependent change in the mode density along the branch cut
by applying the logarithmic argument theorem to the Green function of 
the electromagnetic field. Using the contour sketched in Fig.\ref{BranchCut}(left),
it is possible to show that Eq.\eqref{eq:Eddy} can be written as 
%
%
\begin{equation}
\label{eq:zero-final}
E_{\rm eddy} = \int_{0}^{\infty}\!\frac{{\rm d}\xi}{\pi}
\sum_{p,\mathbf{k}}\ 
\partial_{\xi}\Big(
	\frac{\hbar\xi}{2\pi}\ln \frac{\xi}{\Lambda} 
\Big)
\im{ \ln \left[1-r_{p}^{2}(-\imath \xi-0^{+})e^{-2\kappa L}\right]},
\end{equation}
with $\kappa=\sqrt{\xi^{2}+k^{2}}$ and $r_{p}$ the reflection coefficient of 
the mirrors in polarization $p={\rm TE,TM}$.  This gives rise to a repulsive Casimir
force (Fig.1 of Ref.~\refcite{Intravaia09}),
provided $\Lambda$ is sufficiently large, e.g., $\Lambda \ge \gamma$. 

The structure of Eq.\eqref{eq:zero-final} 
allows for an immediate translation to the high-temperature (classical) limit. 
Replace the zero-point energy with the classical free energy per mode,
$k_{B}T \ln  (\hbar \xi / k_B T)$, and get
\begin{equation}
\label{eq:high-temp}
\mathcal{F}_{\rm eddy}\approx 
- \int_{0}^{\infty}\frac{d\xi}{\pi}\sum_{p,\mathbf{k}}\ 
\frac{k_{B}T}{\xi} \im{ \ln \left[1-r_{p}^{2}(-\imath \xi-0^{+})e^{-2\kappa L}\right]},
\end{equation}
(A more rigorous proof follows from the representation for the free energy 
given in Ref. \refcite{Intravaia09}.) 
Eq. \eqref{eq:high-temp} is thus the result of the logarithmic argument theorem 
applied to the high-temperature limit of the free energy. Now the contour around
the eddy current continuum can also be interpreted as a contour encircling
the whole complex plane, i.e., the surface plasmon and propagating modes
[Fig. \ref{ChangePath}(right)].  
This is particularly interesting in the TE-polarization because there are no surface
plasmons, and 
the residue at $\omega = 0$ vanishes [$r_{\rm TE}^{2}(\omega \to 0) = 0$].
This means that eddy currents and propagating modes give, up to a sign, the
same Casimir energy at high temperature (or large distance). Since propagating
modes are only slightly affected by conduction on the metal (i.e., they behave
similarly in the Drude and plasma models), we find the simple relation
\begin{equation}
\label{eq:high-energy-diff}
\mathcal{F}^{\rm TE}_{\rm eddy} \approx -
\mathcal{F}^{\rm TE}_{\rm C}( {\rm pl.m.} ) 
, \qquad \gamma/\omega_{p}\ll 1
\end{equation}
where $\mathcal{F}^{\rm TE}_{\rm C}( {\rm pl.m.} )$ is the Casimir free 
energy at high temperature calculated within the plasma 
model\cite{Klimchitskaya09}. In the Drude model, the two contributions are
present and cancel each other when they are both in the high-temperature
regime (which happens at different distances, see Fig.4 of 
Ref.\refcite{Intravaia09}).

A different scenario occurs in the TM-polarization. The residue at $\omega = 0$
does not vanish and corresponds exactly to the high-temperature limit of the plasma model.\cite{Klimchitskaya09} 
Indeed, we have checked that eddy currents give only a very 
small contribution.

\section{Conclusions}

Using a mode-summation approach, 
we have isolated and analyzed the contribution
of two classes of modes to the Casimir effect, allowing for complex 
eigenfrequencies of the electromagnetic field. A previous result for the 
short-distance limit between good conductors\cite{Henkel04} 
has been generalized to any conductivity and distance by considering
coupled surface plasmonic modes (for the lossless case, 
see Refs.\refcite{Intravaia05, Intravaia07}).
We also considered eddy currents which are overdamped or diffusive
modes in the bulk of a Drude metal, and 
showed that they contribute a repulsive Casimir interaction, in agreement 
with Ref.\refcite{Intravaia09}. At high temperature and for a good conductor,
we found in a simple way that
their free energy in the TE-polarization differs only slightly from the Casimir 
free energy within a dissipationless description (the plasma model), but is of
the opposite sign.
In this way, eddy currents nearly cancel out the attractive Casimir interaction 
from propagating modes. This explains the
strong difference between the Drude and plasma models for the temperature
correction of the electromagnetic Casimir effect\cite{Klimchitskaya09}.


\smallskip

We thank H. Haakh for a critical reading
and acknowledge financial support by the European Science Foundation 
within the activity `New Trends and Applications of the Casimir 
Effect' (www.casimir-network.com). F.I.\  acknowledges financial support by the Alexander von Humboldt Foundation.

\end{document}